\documentclass[%
 reprint,
superscriptaddress,
 amsmath,amssymb,
 aps,
pre,
floatfix,
]{revtex4-2}

\usepackage[caption=false]{subfig}
\usepackage{graphicx}
\usepackage{dcolumn}
\usepackage{bm}
\usepackage{hyperref}
\usepackage{xcolor}
\usepackage[normalem]{ulem}
\definecolor{ra}{rgb}{0.8, 0.0, 0.0}

\begin{document}

\preprint{APS/123-QED}
\title{Mapping distinct phase transitions to a neural network}
\author{Dimitrios Bachtis}
\email{dimitrios.bachtis@swansea.ac.uk}
\affiliation{Department of Mathematics,  Swansea University, Bay Campus, SA1 8EN, Swansea, Wales, UK}
\author{Gert Aarts}
\email{g.aarts@swansea.ac.uk}
\affiliation{Department of Physics, Swansea University, Singleton Campus, SA2 8PP, Swansea, Wales, UK}
\author{Biagio Lucini}
\email{b.lucini@swansea.ac.uk}
\affiliation{Department of Mathematics,  Swansea University, Bay Campus, SA1 8EN, Swansea, Wales, UK}%
\affiliation{Swansea Academy of Advanced Computing, Swansea University, Bay Campus, SA1 8EN, Swansea, Wales, UK}

\include{ms.bib}

\date{July 7, 2020}

\begin{abstract}
We demonstrate, by means of a convolutional neural network, that the features learned in the two-dimensional Ising model are sufficiently universal to predict the structure of symmetry-breaking phase transitions in considered systems irrespective of the universality class, order, and the presence of discrete or continuous degrees of freedom. No prior knowledge about the existence of a phase transition is required in the target system and its entire parameter space can be scanned with multiple histogram reweighting to discover one. We establish our approach in q-state Potts models and perform a calculation for the critical coupling and the critical exponents of the $\phi^{4}$ scalar field theory using quantities derived from the neural network implementation. We view the machine learning algorithm as a mapping that associates each configuration across different systems to its corresponding phase and elaborate on implications for the discovery of unknown phase transitions.
\end{abstract}

\maketitle

\section{\label{sec:level1}Introduction}
Deep learning \citep{GoodBengCour16} is a category of machine learning algorithms that has recently gained significant importance in the physical sciences. Applications of neural networks have emerged in research fields such as particle physics and cosmology, condensed matter physics, quantum computation and physical chemistry. For a recent review see Ref.~\citep{Carleo_2019}. 

Within these developments machine learning has critically influenced the domain of statistical mechanics, particularly in the study of phase transitions \citep{Carrasquilla2017,vanNieuwenburg2017}. A wide range of machine learning techniques, including neural networks  \citep{Broecker2017,PhysRevX.7.031038,doi:10.7566/JPSJ.86.063001,PhysRevB.94.165134,PhysRevB.97.174435,PhysRevE.100.052312,PhysRevLett.121.245701,PhysRevB.98.060301,PhysRevLett.120.257204,alex2019unsupervised,chernodub2020topological,PhysRevD.100.011501,PhysRevE.101.023304}, diffusion maps \citep{Rodriguez-Nieva2019}, support vector machines \citep{PhysRevB.96.205146,GIANNETTI2019114639,PhysRevB.99.060404,PhysRevB.99.104410} and principal component analysis \citep{PhysRevB.94.195105,PhysRevE.95.062122, PhysRevB.96.144432,PhysRevB.96.195138, PhysRevB.99.054208} have been implemented to study equilibrium and non-equilibrium systems. Transferable features have also been explored in phase transitions, including modified models through a change of lattice topology \citep{Carrasquilla2017} or form of interaction \citep{PhysRevB.100.045129}, in Potts models with a varying odd number of states \citep{Shiina2020}, in the Hubbard model\citep{PhysRevX.7.031038}, in fermions \citep{Broecker2017}, in the neural network-quantum states ansatz \citep{PhysRevE.101.053301,Carleo_2017manybody} and in adversarial domain adaptation \citep{PhysRevB.97.134109}. Among these, recent studies based on transferable features have mostly focused on predicting the critical temperature exclusively.

A machine learning algorithm can be implemented, based on a set of labeled training data, to complete a classification task such as the separation of phases in a statistical system. This is almost universally conducted under the assumption that the training and prediction data belong in identical probability distributions, and have been acquired from the same feature space. A change in feature space could arise by attempting to separate phases in a different system. However such an attempt would require the algorithm to be re-built, due to the difference in systems, in order to solve, in essence, a new classification task. Transfer learning is a framework within the research field of machine learning introduced to address precisely this problem \citep{Pan10asurvey}. Specifically, it enables the transfer of knowledge from a solved machine learning task to a new one, that shares certain similarities with the original. 

In this paper, we propose transfer learning as a means to discover and study unknown phase transitions. In particular, we explore if the features learned by a convolutional neural network on the two-dimensional Ising model are sufficiently universal to predict the structure of phase transitions in systems irrespective of the universality class, order, and discrete or continuous degrees of freedom. We further explore the deeper layers of the neural network architecture for universal features and discuss how the neural network acts as a mapping, which associates each configuration, across different systems under consideration, to its corresponding phase. It is then feasible to predict phases across a wide range of systems, governed by distinct Hamiltonians and symmetries, without introducing prior knowledge about the presence of a phase transition in them.

\begin{figure*}
\includegraphics[width=16.2cm]{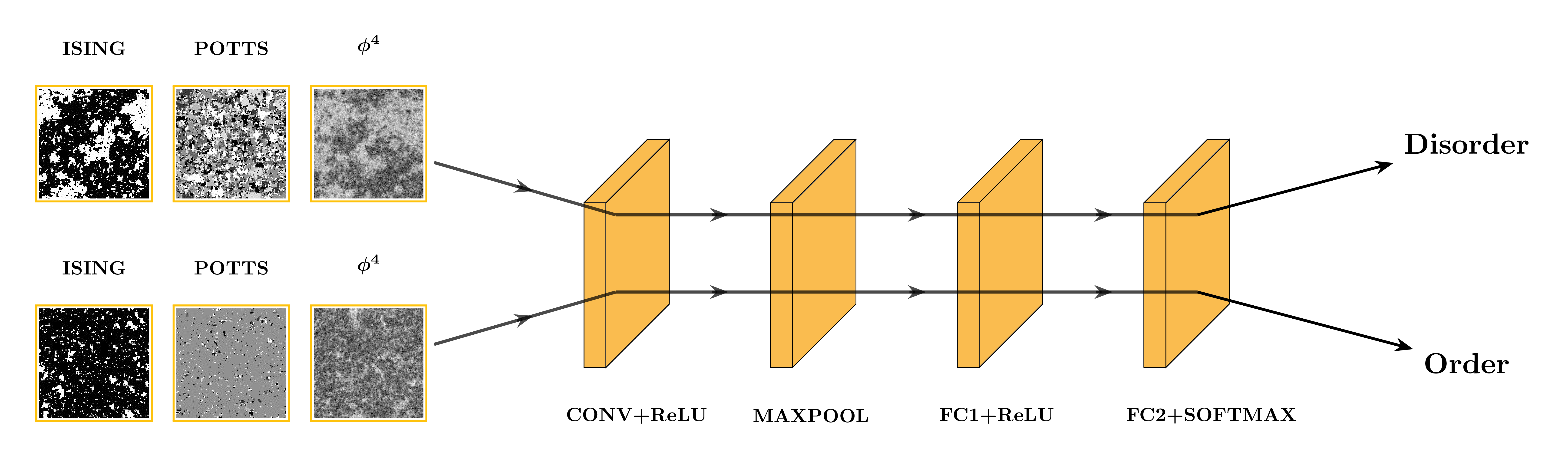}
\caption{\label{fig:conv2d}The 2D Ising-trained convolutional neural network (see App.~\ref{app:cnndetails}).  Configurations are successfully classified as belonging to the disordered or ordered phase, irrespective of the system.}
\end{figure*}

To study unknown phase transitions, we treat the predictive function of the Ising-trained neural network as an observable in the target system, and propose the extension of previous work on reweighting for machine learning \citep{bachtis2020extending}, to the multiple histogram method \citep{PhysRevLett.63.1195,PhysRevLett.61.2635}. This technique enables the accurate definition of the target system's critical region by interpolating the neural network function in its entire parameter space. Given the knowledge of an effective order parameter in a target system through transfer learning, we train a randomly initialized neural network to perform a finite-size scaling analysis based on histogram-reweighted quantities derived from the machine learning algorithm. This results in an accurate calculation of multiple critical exponents and the critical coupling.

To establish our approach, we apply the Ising-trained neural network to q-state Potts models, systems that manifest first-order or second-order phase transitions depending on their number of states, and the $\phi^{4}$ scalar field theory.  Without using knowledge about the presence of a phase transition in the target system, we employ transfer learning to reconstruct an effective order parameter therein, define rigidly the boundaries of its critical region, and perform a finite size scaling analysis on machine learning derived quantities to determine its universality class.
\section{\label{sec:level2}Transfer Learning}

\begin{figure*}[t]
\subfloat[\label{fig:mhrew}]{%
  \includegraphics[width=8.6cm]{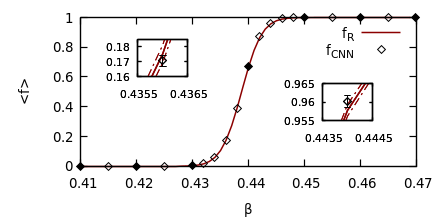}%
}\hfill
\subfloat[\label{fig:phi}]{%
  \includegraphics[width=8.6cm]{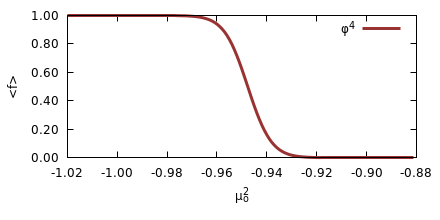}%
}\hfill
\subfloat[\label{fig:potts}]{%
  \includegraphics[width=16.2cm]{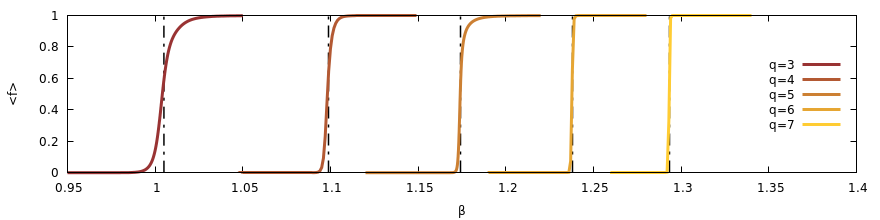}%
}
\caption{Multiple histogram reweighting of the predictive function versus coupling of the two-dimensional (a) Ising model, (b) $\phi^{4}$ scalar field theory and (c) $q=3,\ldots,7$ (left to right) Potts models,  for lattice size $L=128$. The reweighted predictive function is depicted by the line. Its statistical errors are smaller than its width, excluding the insets in Fig.~(a), where errors are portrayed by dashed lines. The results for all figures have been obtained based on a convolutional neural network trained exclusively on the Ising model. In Fig.~(a) filled points correspond to the combined Monte Carlo datasets used to conduct reweighting and empty points are calculations of the predictive function on independent Monte Carlo datasets, added for comparison. In Fig.~(c) vertical dashed lines have been added for the values of critical inverse temperature of Potts models, calculated through Eq.~\ref{eq:crittemppotts}.}\label{fig:a}
\end{figure*}

We consider a domain $\Delta$ which is comprised of a feature space $\Omega$ and a marginal probability distribution $P(X)$, where $X=\lbrace x_{i},\ldots,x_{N} \rbrace \in \Omega$ is a learning sample:
\begin{equation}\label{eq:domain}
\Delta= \lbrace \Omega, P(X) \rbrace.
\end{equation}
For the case of the Ising model, the feature space $\Omega$ contains all possible configurations of the system, $X$ is a subset of $\Omega$, comprised of a finite number of configurations that have been drawn from selected inverse temperatures, and $x_{i}$ corresponds to a specific configuration.

Within a domain $\Delta$, we define a task $T$, with a label space $Y$ and a predictive function $f(\cdot)$ which is learned during the optimization of the machine learning algorithm on the training data:
\begin{equation}
T=\lbrace Y, f(\cdot) \rbrace.
\end{equation}
In particular, for a phase identification task in the Ising model, a label $y_{i} \in Y$, denoted by zero or one, signifies the disordered or the ordered phase and the predictive function $f(\cdot)$, which is equivalent to the conditional probability distribution $p(y | x)$, associates each configuration $x$, given as input to the algorithm, to its corresponding phase.

Given the knowledge acquired on a source domain $\Delta_{s}$ and learning task $T_{s}$, transfer learning can be utilized for a target domain $\Delta_{t}$ and learning task $T_{t}$, to enhance the predictive function $f(\cdot)_{t}$, when $\Delta_{s} \neq \Delta_{t}$ or $T_{s} \neq T_{t}$ \citep{Pan10asurvey}. The condition $\Delta_{s} \neq \Delta_{t}$ implies, based on Eq.~\ref{eq:domain}, that the feature spaces  or the marginal probability distributions might be different: $\Omega_{s} \neq \Omega_{t}$, $P_{s}(X) \neq P_{t}(X)$. An example of domain adaptation concerns employing transfer learning to relocate from a source domain of a two-dimensional binary system, such as the Ising model, to a different system in search for a phase transition that separates a disordered from an ordered phase.  

\section{\label{sec:level3}Discovering Phase Transitions}

We employ transfer learning to predict the phase diagram of two-dimensional q-state Potts models and the $\phi^4$ scalar field theory using an Ising-trained convolutional neural network and multiple histogram reweighting. In addition we explore if the neural network accurately classifies distinct phase transitions due to the presence of learned universal features in deeper layers.  

The configurations of the Ising and the Potts models are obtained using Markov chain Monte Carlo simulations with the Wolff algorithm \citep{PhysRevLett.62.361}. The scalar field theory is simulated with the Metropolis algorithm followed by a sweep of the Wolff algorithm \citep{PhysRevD.58.076003,PhysRevLett.62.1087} (see App.~\ref{app:allmodels} for details of the models). The convolutional neural network (see Fig.~\ref{fig:conv2d}) is trained on the Ising model, where configurations have binary degrees of freedom, mapped to $-1$ and $1$.   Training of the neural network is conducted for $\beta \leq 0.41$ and $\beta \geq 0.47$ in the disordered and ordered phases, respectively  (see App.~\ref{app:cnndetails} for details of the neural network). To be consistent with the same range when conducting transfer learning, configurations from Potts models are mapped to unique positive and negative numbers between $-1$ and $1$. By choosing unique values for the states of the Potts model the physics of the Hamiltonian, which includes a delta function, is retained. When the number of states is odd, the remaining value is chosen arbitrarily as positive or negative. The degrees of freedom for the $\phi^{4}$ scalar field theory lie within $-\infty < \phi < \infty$ and no normalization is conducted.

Once the convolutional neural network is trained on the source domain $\Delta_{s}$, to classify the ordered and disordered phases of the Ising model, its predictive function $f_{s}(\cdot)$, which is interpreted here as the probability of being in the ordered phase, is employed to predict the phase of a configuration $x$. The application of the predictive function to an importance-sampled configuration $x$, converts $f_{s}(x)$ into an observable with an attached Boltzmann weight, enabling its extrapolation with reweighting in the statistical system's parameter space \citep{bachtis2020extending}. Here, we propose the extension of reweighting to the multiple histogram method (see App.~\ref{app:mhderiv} for a derivation). This technique combines an arbitrary number of simulated datasets and enables the estimation of the partition function, and therefore of the predictive function, in the system's entire parameter space \citep{PhysRevLett.63.1195}.

\begin{figure*}[t]
\includegraphics[width=16.2cm]{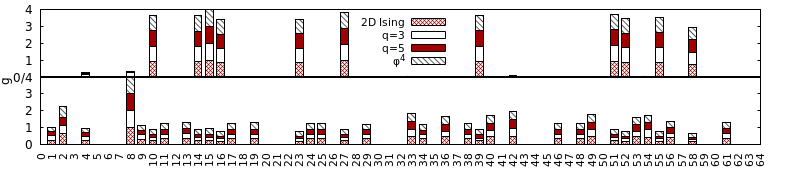}
\caption{\label{fig:feat}Mean activations $g$ versus the 64 latent variables in the fully connected layer (FC1) of the 2D Ising-trained convolutional neural network. Results are from the two-dimensional Ising model, q=3 and q=5 Potts model, and the $\phi^4$ field theory for configurations belonging in the disordered (top) and ordered phase (bottom). The maximum values have been normalized to one and the results have been rescaled accordingly and stacked vertically for better representation. }
\end{figure*}

The results can be seen in Fig.~\ref{fig:mhrew}, where the predictive function, depicted by the line, is obtained in the visible range using multiple histogram reweighting on a combination of Monte Carlo datasets. We recall that the CNN is trained on the intervals $\beta \leq 0.41$ and $\beta \geq 0.47$. The results are compared with calculations of the neural network on independent Monte Carlo simulations,  which lie within statistical errors. The predictive function acts as an effective order parameter and the use of multiple histogram reweighting advances previous work by enabling the estimation of the predictive function over extended ranges in the system's parameter space.

The predictive function $f_{s}(\cdot)$ was learned on configurations of the Ising model, but the classification knowledge can be adapted to other domains by applying it to configurations $x'$ of a different system. The predictive function $f_{s}(x')$ then becomes an observable in the target system, where multiple histogram reweighting can be employed to estimate it in the target parameter space. The results for Potts models can be seen in Fig.~\ref{fig:potts} where we recall that the phase transition is second-order for $q=3,4$ and first-order for $q=5,6,7$ \citep{Baxter}. The dashed vertical lines indicate the positions of the analytically determined critical coupling, $\beta_{c} = \ln(1+\sqrt q)$. Good agreement between the results obtained by transfer learning and the exact ones can already be observed on the volume shown here ($L=128$). The results for the $\phi^{4}$ scalar field theory, a system with continuous degrees of freedom, can be seen in Fig.~\ref{fig:phi}, where we have fixed the dimensionless quartic coupling $\lambda_{L}=0.7$ and reweighted based on the values of $\mu^{2}$,  the dimensionless mass parameter (see App~\ref{app:allmodels}). We infer that the system is in the broken (ordered) phase for large and negative $\mu^2$ and in the symmetric (disordered) phase for smaller $\mu^2$ values.  In Sec.~\ref{sec:stud} we analyse the critical properties of the scalar field theory further using a finite-size scaling analysis. 

We note that the neural network successfully differentiates between ordered and disordered phases, irrespective of the system, and despite changes in discrete or continuous degrees of freedom, the universality class and the order of the phase transition. Consequently, the predictive function $f_{s}(\cdot)$ learned by a convolutional neural network on the two-dimensional Ising model is capable to reconstruct effective order parameters in more complicated systems.

To gain further insights about the capability of the neural network to predict phases across different systems, we consider that the fundamental constituents of a trained neural network are the sets of variational parameters, comprised of the weights and biases at each layer of its architecture. These are optimized during the training process for a specific machine learning task $T$. The variational parameters eventually converge to certain values which encode the solution to the particular problem under consideration. In our approach they have been tuned to accurately separate the disordered and ordered phases of the two-dimensional Ising model based on a set of labeled Ising configurations. It is well known that within the first layers of a neural network architecture the variational parameters correspond to learned universal features \citep{NIPS2014_5347}. This form of universality is expected to eventually diminish towards deeper layers of the neural network architecture where the features are anticipated to transition from universal to specific in relation to the machine learning task $T$.

Considering that the 2D Ising-trained neural network successfully classifies phases across different systems by utilizing an identical predictive function $f_{s}(\cdot)$, we expect universal features to extend further in deeper layers. The activation function of a variable in an intermediate layer of the neural network architecture acts as a transformation that maps a certain input to an output representation. For a feature to be deemed universal in relation to certain inputs, the corresponding activation function should produce identical output representations. We therefore present as input to the neural network configurations from different systems, and calculate their mean activations in order to diminish statistical errors during classification. The results are depicted in Fig.~\ref{fig:feat}, where the activations have been drawn for the 64 variables of the first fully connected layer. We observe spikes for values of certain activations, which are consistent for configurations from the ordered or disordered phase, irrespective of the system. The features learned by the Ising-trained convolutional neural network successfully map each configuration, across systems under consideration, to its associated phase.

\section{\label{sec:stud}Studying unknown phase transitions}
The preceding results designate that a neural network trained on a prototypical system, such as the Ising model which manifests a second-order phase transition, can serve as a tool to discover phase transitions in more complicated systems. This pursuit is greatly enhanced with the use of multiple histogram reweighting, where the entire parameter space can be examined for the discovery of a phase transition.  By obtaining the knowledge of an effective order parameter through the use of transfer learning in a target system, the boundaries of its critical region can then be accurately defined. 

 As an example, for the $\phi^{4}$ field theory with fixed $\lambda_{L}=0.7$, we presume based on Fig.~\ref{fig:phi}, that a conservative choice for the boundaries of the ordered and disordered phases lie on values of bare mass $\mu^{2} \leq -1.0 $ and $\mu^{2} \geq -0.90$, respectively. We note that one could employ the neural network that produced the results of Fig.~\ref{fig:phi} to study the infinite-volume limit of the target system. However, since target systems might differ from the source system in fundamental aspects such as the order of the phase transition, the continuous degrees of freedom or the underlying symmetries, we treat the preceding results from transfer learning as qualitative indications. To obtain quantitative results, we train a randomly initialized neural network on the $\phi^{4}$ scalar field theory to study the infinite volume limit. We sample configurations which are labeled as belonging to each phase based on Fig.~\ref{fig:phi}. For the randomly initialized neural network, which is trained using the same setup (see App.~\ref{app:cnndetails}), the source domain is the one defined by the configurations of $\phi^{4}$ scalar field theory and the machine learning task $T$ is the separation of the two phases, which are discovered through transfer learning. 
 
 \begin{figure}[t]
\includegraphics[width=8.6cm]{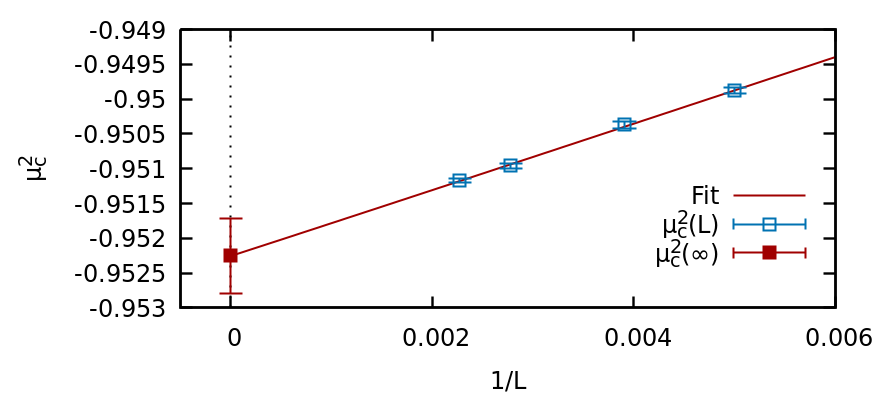}
\caption{\label{fig:fss} Pseudo-critical $\mu_{c}^{2}$ versus inverse lattice size.}
\end{figure}
\begin{table}[b]
\caption{\label{tab:data}
Pseudo-critical points $\mu_{c}^{2}(L)$ for fixed $\lambda_{L}=0.7$ and maxima of the predictive function $\delta f_{max}$ for various lattice sizes $L$ of the $\phi^{4}$ scalar field theory.}
\begin{ruledtabular}
\begin{tabular}{ccc}
$L$ &$\mu_{c}^{2}(L)$&$\delta f_{max}$ \\
\hline
200 & -0.94988(4) & 8239(50) \\
256 & -0.95037(5) & 12915(56) \\
360 & -0.95096(4) & 22348(138) \\
440 & -0.95117(3) & 34710(211)
\end{tabular}
\end{ruledtabular}
\end{table}

 \begin{figure}[t]
\includegraphics[width=8.6cm]{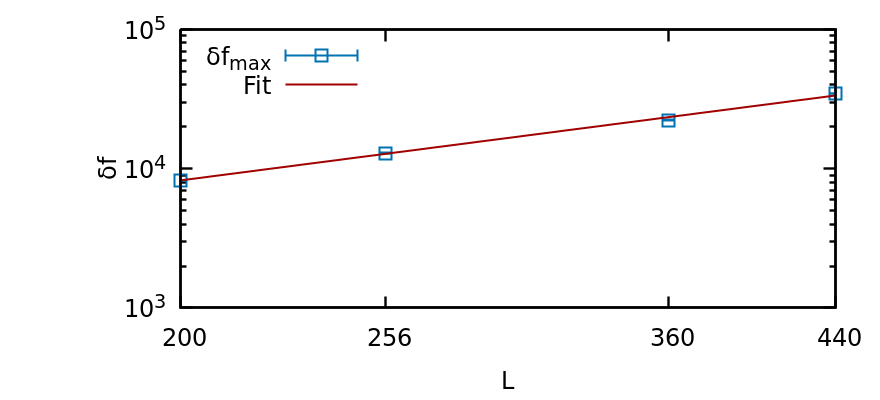}
\caption{\label{fig:fssg}  Fluctuations of the predictive function versus lattice size on double logarithmic scale.}
\end{figure}

We proceed by performing a calculation of the critical $\mu_{c}^{2}$ and the critical exponents of the $\phi^{4}$ scalar field theory under a finite size scaling assumption, by relying on quantities derived from the neural network implementation and their reweighted extrapolations \citep{bachtis2020extending}. Specifically, we associate pseudo-critical points based on the maxima of the fluctuations of the predictive function:
\begin{equation}
\delta f(\cdot)= V(\langle f^{2} \rangle-\langle f \rangle^2),
\end{equation}
where $V$ is the volume of the system. The maxima (see Table~\ref{tab:data}) are obtained by reweighting based on one Monte Carlo dataset within the critical region. We note that including measurements from additional simulations in reweighting always leads to a reduction in statistical errors \citep{PhysRevLett.63.1195}. We then calculate the correlation length critical exponent and critical $\mu_{c}^{2}$ through equation:
\begin{equation}
\Bigg| \frac{\mu^{2}_{c}(L)-\mu^{2}_{c}}{\mu^{2}_{c}} \Bigg| \sim \xi^{-\frac{1}{\nu}} \sim L^{-\frac{1}{\nu}},
\end{equation}
and the magnetic susceptibility exponent through equation:
\begin{equation}
\delta f \sim L^{\frac{\gamma}{\nu}}.
\end{equation}
 \begin{table}[t]
\begin{ruledtabular}
\begin{tabular}{ccccc}
 &$\mu_{c}^{2}$&$\nu$ &$\gamma/\nu$ \\
\hline
CNN+Reweighting &  -0.95225(54)  & 0.99(34) & 1.78(7) \\
\end{tabular}
\end{ruledtabular}
\caption{\label{tab:table2}
Critical $\mu_{c}^{2}$ for fixed $\lambda_{L}=0.7$ and critical exponents of the $\phi^4$ scalar field theory.}
\end{table}
The results are presented in Figs.~\ref{fig:fss} and~\ref{fig:fssg} and in Table~\ref{tab:table2}. We note that the value of the critical $\mu_{c}^{2}$ for fixed $\lambda_{L}=0.7$ is within statistical errors from calculations of conventional phase transition indicators, such as the susceptibility or the Binder cumulant, in Refs.~\citep{PhysRevD.58.076003,PhysRevD.79.056008}. For the values of $\lambda_{L}$ and $\mu^{2}$, we anticipate the phase transition to be in the universality class of the two-dimensional Ising model, as evidenced by the calculation of the critical exponents (see App.~\ref{app:allmodels}). In the error analysis (see App.~\ref{binning}), only statistical errors from predictions on a finite Monte Carlo dataset were considered. Better accuracy can be achieved by including calculations from additional lattice sizes.
\section{Conclusions}
In this paper we employed transfer learning to discover and study phase transitions from ordered to disordered phases in two-dimensional statistical systems. By employing an Ising-trained convolutional neural network and multi-histogram reweighting, we reconstructed effective order parameters in q-state Potts models and the $\phi^{4}$ scalar field theory, without introducing prior knowledge about the presence of a phase transition in those target systems. In addition we viewed the neural network as a mapping that associates each configuration, across different systems, to its corresponding phase and uncovered universal features learned on deeper layers of the neural network architecture. Furthermore we utilized transfer learning to define the boundaries of the critical region in the $\phi^{4}$ scalar field theory. We finally conducted a finite size scaling analysis to calculate multiple critical exponents and the critical $\mu_{c}^{2}$ constant using quantities derived solely from the neural network algorithm. We are not aware of any prior work that employs machine learning to calculate critical exponents for a quantum field theory.

The discovery and study of an unknown phase transition can be completed in four steps:
\begin{itemize}
\item A neural network is trained on labeled configurations of an original system to learn a predictive function that separates a number of phases.
\item The predictive function is then applied on unlabeled configurations of a target system, to recognize if they belong to any of the learned phases. The entire parameter space can be inspected with multiple histogram reweighting to uncover the target system's critical region.
\item Given this knowledge a second neural network is trained on configurations of the target system that have been labelled in compliance with the recognized phases in the previous step.
\item The infinite volume limit is studied, using reweighting on derived quantities from the second neural network implementation to determine the universality class of the phase transition in the target system.
\end{itemize}
Transfer learning enables the use of simplistic systems to study complicated models with partially known behaviour. Combined with multi-histogram reweighting, a technique that is able to scan the entire parameter space by reconstructing effective order parameters therein, it permits the discovery of unknown phase transitions. Transfer learning is employed successfully to predict the phase structure of target systems irrespective of the universality class, order, and discrete or continuous degrees of freedom, and can therefore be employed to uncover phase transitions in seemingly unrelated systems. Finally one could combine multiple transfer learning implementations, obtained by training neural networks on models with different phase transitions, to predict the structure of an intricate, unknown phase transition in a target system. 

Note added: During completion of this manuscript, Refs.~\citep{yau2020generalizability,mendessantos2020unsupervised} appeared on the arXiv. Ref.~\citep{yau2020generalizability} employs transfer learning for Potts models, and Ref.~\citep{mendessantos2020unsupervised} explores universal critical behaviour of phase transitions based on the intrinsic dimension in the data space.
\section{\label{sec:level5}Acknowledgements}
The authors received funding from the European Research Council (ERC) under the European Union's Horizon 2020 research and innovation programme under grant agreement No 813942. The work of GA and BL has been supported in part by the STFC Consolidated Grant ST/P00055X/1. The work of BL is further supported in part by the Royal Society Wolfson Research Merit Award WM170010. Numerical simulations have been performed on the Swansea SUNBIRD system. This  system is part of the Supercomputing Wales project, which is part-funded by the European Regional Development Fund (ERDF) via Welsh Government. We thank COST Action CA15213 THOR for support.
\appendix
\section{\label{app:allmodels}Ising, Potts and the $\phi^{4}$ field theory}
We consider the q-state Potts model on a square lattice, with a Hamiltonian:
\begin{equation}
H_{P}=-J \sum_{\langle ij \rangle} \delta (\sigma_{i},\sigma_{j}),
\end{equation}
where $J$ denotes the coupling constant, which we set to one, $\langle ij \rangle$ is a sum over nearest neighbour interactions, $\delta (\sigma_{i},\sigma_{j})$ is the Kronecker delta and $\sigma_{i}$ is a spin at lattice site $i$ that can take the values $1,\ldots,q$. The 2D Potts model has a second-order phase transition for $q \leq 4$ and a first-order phase transition for $q \geq 5$ \citep{Baxter}. For $q=2$, $\sigma \in \lbrace -1,1 \rbrace$, $J_{\rm Ising}=J/2$, the Potts model reduces to the Ising model. The inverse critical temperature of the 2D q-state Potts model is given by:
\begin{equation}\label{eq:crittemppotts}
\beta_{c}^{\rm{Potts}}= \ln(1+\sqrt{q}),
\end{equation} 
where for $q=2$, $\beta^{\rm{Ising}}_{c}=\beta_{c}^{\rm{Potts}}/2$.

We consider the $\phi^{4}$ scalar field theory in two dimensions, described by the Euclidean Lagrangian:

\begin{equation}
\mathcal{L}_{E}= \frac{1}{2} (\nabla \phi)^{2} + \frac{1}{2} \mu_{o}^{2} \phi^{2} + \frac{\lambda}{4} \phi^{4}.
\end{equation}

We discretize the $\phi^{4}$ scalar field theory on a square lattice with lattice spacing $a$: 
\begin{equation} \nonumber
S_{E}= \sum_{n} \Bigg[   \frac{1}{2} \sum_{\nu=1}^{d=2} (\phi_{n+e_{\nu}}-\phi_{n})^{2} + \frac{1}{2} \mu^{2} \phi_{n}^{2} + \frac{1}{4} \lambda_{L} \phi_{n}^{4} \Bigg],
\end{equation}
where the dimensionless parameters $\mu^{2}= a^{2} \mu_{0}^{2}$ and $\lambda_{L}=a^{2} \lambda$ are the (bare) mass squared and coupling constant. By fixing the value of $\lambda_{L}$ while varying $\mu^{2}$ we cross a second-order phase transition in the $(\lambda_{L},\mu^{2})$ plane \citep{PhysRevD.58.076003}. When $\lambda_{L} \rightarrow \infty$ and $\mu^{2} \rightarrow -\infty$ the system is expected to be in the same universality class as the 2D Ising model, whereas for $\lambda_{L}$ fixed and $\mu \rightarrow 0$ the emerging critical behaviour is Gaussian \citep{Milchev1986}.
\section{\label{app:cnndetails}Convolutional Neural Network}
The neural network architecture, implemented with TensorFlow and the Keras library, is comprised of a convolutional layer with 64 filters, of size $2 \times 2$ and a stride of $s=2$.  The result is then forwarded to a max-pooling layer of size $2 \times 2$ and subsequently to a fully-connected layer (FC1) with 64 latent variables. The non-linear function is chosen as a rectified linear unit $g(x)=\max(0,x)$. The output layer consists of a fully connected layer (FC2) with a softmax activation function and the training is conducted with the Adam algorithm and a mini batch size of 12. The architecture was optimized for the Ising model based on the training and validation loss.

The neural network is trained on 1000 uncorrelated configurations per each parameter choice. Specifically for the case of the Ising model the range of inverse temperatures is $0.32,\dots,0.41$ and $0.47,\ldots,0.56$  in the disordered and the ordered phase respectively, with a step size of $0.01$. The second neural network is trained independently on the $\phi^{4}$ field theory on values of bare mass $-1.09,\ldots,-1.00$ and  $-0.90,\ldots,-0.81$ with the same step size.
\section{\label{app:mhderiv}Multi-Histogram Reweighting}
During a Monte Carlo simulation the probability of generating a state with a lattice action (or energy) $S$ is:
\begin{equation}
p(S)= \rho(S) \frac{\exp[{-\sum_{k} g^{(k)} S^{(k)}}]}{Z} ,
\end{equation}
where $\rho(S)$ the density of states, $Z$ the partition function and $S$ a lattice action  (or Hamiltonian) that separates in terms of a set of parameters $\lbrace g^{(k)} \rbrace$:
\begin{equation}
S= \sum_{k} g^{(k)} S^{(k)}.
\end{equation}
The probability $p(S)$ can be estimated during a Monte Carlo simulation through $p(S)= N(S)/n$ where $n$ is the number of independent measurements and $N(S)$ the histograms of the action. After conducting a series of simulations for a specific set of parameters $\lbrace g_{i}^{(k)} \rbrace$, we arrive at a number of different estimates $\rho_{i}$ of the density of states, given by:
\begin{equation}
\rho_{i}(S)= \frac{N_{i}(S) Z_{i}}{n_{i} \exp\Big[{-\sum_{k} g^{(k)}_{i} S^{(k)}}\Big]}.
\end{equation}
The estimates can be combined through a weighted average to estimate the density of states :
\begin{equation}
\rho(S) =\sum_{i} w_{i} \rho_{i}(S). 
\end{equation}
The values $w_{i}$ are acquired through a minimization of the variance in the density of states, resulting in the equation:
\begin{equation}
\rho(S)= \frac{\sum_{i} N_{i}(S)}{\sum_{j} n_{j} Z_{j}^{-1} \exp\Big[{-\sum_{k} g^{(k)}_{j} S^{(k)}}\Big]},
\end{equation}
where $j$ is the number of available Monte Carlo datasets. For each dataset, simulated on a parameter set $\lbrace g_{m}^{(k)} \rbrace$, the partition function, given by equation:
\begin{equation}
Z_{m}= \sum_{S} \rho(S) \exp[{-\sum_{k} g^{(k)}_{m} S^{(k)}}],
\end{equation}
can be estimated through an iterative scheme \citep{PhysRevLett.63.1195, newmanb99}, by solving:
\begin{equation} \label{partiter}
Z_{m}= \sum_{i,s} \frac{1}{\sum\limits_{j} n_j Z_{j}^{-1} \exp\Big[{\sum\limits_{k} (g^{(k)}_{m}-g^{(k)}_{j}) S_{is}^{(k)}}\Big]}.
\end{equation}
\begin{figure}[b]
\includegraphics[width=8.6cm]{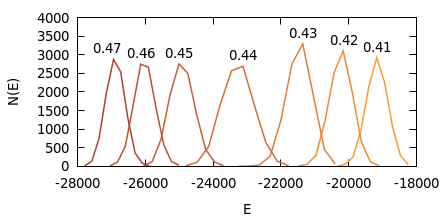}
\caption{\label{fig:mhistograms} Histograms of energies, from Monte Carlo datasets simulated at inverse temperatures $\beta=0.41 \ldots 0.47$ of the 2D Ising model for $L=128$. Overlapping histograms enable the interpolation over the entire inverse temperature range with multiple histogram reweighting.}
\end{figure}
After convergence, the partition function $Z_{l}$ for an interpolated parameter set $\lbrace g_{l}^{(k)} \rbrace$ can be acquired by one iteration of Eq.~\ref{partiter}. The expectation value of an arbitrary observable $\langle O \rangle_{l}$ for a set of parameters $\lbrace g_{l}^{(k)} \rbrace$ is then calculated using equation:
\begin{equation} \label{eq:observable}
\langle O \rangle_{l} = \frac{1}{Z_{l}} \sum_{i,s} \frac{O_{is}}{\sum\limits_{j} n_{j} Z_{j}^{-1} \exp\Big[{\sum\limits_{k} (g^{(k)}_{l}-g^{(k)}_{j}) S_{is}^{(k)}}\Big] },
\end{equation}
where $s$ is the number of states sampled at $i_{th}$ simulation.  By carefully conducting simulations in order to obtain overlapping histograms between parameter values (e.g. see Fig.~\ref{fig:mhistograms}) one can calculate the partition function and arbitrary observables of interest in the entire parameter space using Eqs.~\ref{partiter} and~\ref{eq:observable}.
\section{\label{binning}Binning Error Analysis}
The error analysis is conducted with a binning approach. A Monte Carlo dataset, comprised of 10000 uncorrelated measurements, is separated into $n_{b}=10$ groups of 1000 measurements. Multi-histogram reweighting is conducted $n_{b}$ times for extrapolated sets of parameters using the measurements of each group in each available simulation. The standard deviation for the expectation value at each extrapolated parameter is given through equation:
\begin{equation}
\sigma= \sqrt{\frac{1}{n_{b}-1} (\overline{O^{2}}-\overline{O}^{2})}.
\end{equation}

\bibliography{ms}
\end{document}